**Ante, L., Wazinski, F.-P., & Saggu, A. (2026). Research Streams in Biodiversity Finance: A Bibliometric Analysis and Research Agenda. Finance Research Open, 100123.**

# Research Streams in Biodiversity Finance: A Bibliometric Analysis and Research Agenda

**Abstract:** Biodiversity loss is accelerating at an unprecedented pace, threatening ecosystem stability, economic resilience, and human well-being, with billions required to reverse current trends. Against this backdrop, biodiversity finance has emerged as a rapidly expanding but highly fragmented field spanning ecology, economics, finance, accounting, and policy. Biodiversity finance supports the conservation, sustainable use, and restoration of biodiversity through targeted financial mechanisms. However, it remains an emerging, complex, and fragmented field, with the majority of relevant knowledge being produced in non-finance journals. This study employs quantitative bibliometric analysis to examine a corpus of 189,456 references underlying 3,998 articles related to biodiversity and finance. The analysis identifies eight primary research streams within the field that concern (1) strategic and financial approaches in global biodiversity conservation, (2) the impact and implementation of payments for environmental services (PES) in developing countries, (3) neoliberal influences and implications in environmental conservation, (4) biodiversity offsets and conservation, (5) ecosystem services and biodiversity, (6) integrating conservation and community interests in biodiversity management, (7) balancing agricultural intensification with biodiversity conservation, and (8) global and corporate biodiversity reporting. The characteristics of each research stream and its prevalent publications are outlined, alongside an analysis of their temporal evolution and the degree of information exchange among the research streams. The findings provide a structured map of the intellectual architecture of biodiversity finance, document pronounced silos between economically-oriented and critical/political-economy research streams, and translate these patterns into a focused research agenda and implications for policymakers, financial institutions, and corporate actors.

**Keywords:** Biodiversity finance; Environmental finance; Conservation finance; Sustainable development; Ecosystem services

**JEL Codes:** Q56; Q57; Q58; G18; G23

## 1 Introduction

Biodiversity loss, driven by deforestation, habitat destruction, pollution, and climate change, poses severe risks to ecosystems and human well-being. The ongoing decline in biodiversity impacts food security, water supply, health, and economic stability, with the potential to exacerbate poverty and social inequality (Cardinale et al., 2012; Hooper et al., 2012; Reich et al., 2012). Despite global efforts to address this crisis, including commitments such as the Aichi Biodiversity Targets established under the Convention on Biological Diversity (CBD), progress has been slow (Buchanan et al., 2020; Hagerman and Pelai, 2016). The failure to meet these biodiversity targets has highlighted the need for innovative approaches to conservation and sustainable use of biodiversity, with finance being a pivotal element.

Biodiversity finance refers to financial instruments, policies, or investments that aim to promote the conservation and sustainable use of biodiversity (Flammer et al., 2023; Karolyi and Tobin-de la Puente, 2023). With an estimated financing gap of $589 to $824 billion needed annually to halt biodiversity loss by 2030[1] (Deutz et al., 2020), financial markets and investors are integral to achieving global biodiversity goals. Much like climate finance, biodiversity finance faces challenges such as a lack of adequate funding, higher investment risks, or lower direct short-term returns, which disincentivize private capital from flowing into conservation projects (Barbier et al., 2018; Thompson, 2023). Furthermore, the links between biodiversity loss and the financial system are complex. For example, deforestation and ecosystem degradation can result in physical risks for industries dependent on natural resources (Calice et al., 2023), while transitioning to a nature-positive economy may create regulatory and market risks for sectors like agriculture, forestry, and mining (Kousky, 2022; van Rees et al., 2023).

Similar to green finance (Ante, 2024), biodiversity finance encompasses a range of financial tools, including green bonds, biodiversity offset credits, impact investments, and conservation trust funds (Flammer et al., 2023; Karolyi and Tobin-de la Puente, 2023). However, unlike broader environmental finance, biodiversity finance specifically targets nature-based solutions that restore ecosystems and promote sustainable development. These investments deliver environmental benefits but also align with the Sustainable Development Goals (SDGs), particularly SDG 14 (Life Below Water) and SDG 15 (Life on Land). Nonetheless, biodiversity finance remains underdeveloped compared to other areas of sustainable finance, with financial institutions struggling to incorporate biodiversity risks and opportunities into their strategies (Bassen et al., 2024; Nedopil, 2023).

The growing recognition of biodiversity as a critical factor for long-term economic resilience has spurred increased interest in biodiversity finance. Yet despite the field's growing importance, limited research has explored the biodiversity finance literature using informetric methods[2]. Further, the majority of studies on financial aspects of biodiversity are published outside traditional finance journals[3]. This makes it necessary to provide the subject area with an objective, comprehensive basis to reduce search costs. This study aims to fill this gap by conducting a comprehensive bibliometric analysis of the biodiversity finance literature. By applying co-citation analysis, this study identifies key research streams, highlights influential publications, and maps the evolution of biodiversity finance research. The objective is to provide a structured overview of the field, informing both policymakers and academics about the underlying structure, emerging trends, and potential areas for future research.

The analysis is based on a collection of 3,998 articles on biodiversity finance sourced from the Scopus database. Using co-citation data, eight major research streams in biodiversity finance are identified through factor analysis. The study reviews the fit and relevance of publications to these streams, visualizes their temporal development, and analyzes the flow of information across the different research streams. This approach aims to offer a clear, objective picture of biodiversity finance, highlighting key contributions and revealing gaps in the existing literature that could guide future research.

---

[1] Deutz et al. (2020) estimate the required spending on biodiversity conservation between $722 to $967 billion per year, whereas current spending sits between $124 and $143 annually.

[2] To our knowledge, the short bibliometric analysis of Hutchinson and Lucey (2024) represents the sole existing bibliometric study focusing on biodiversity finance. However, the study builds on a comparatively smaller dataset of primary studies. It analyses the status quo of research using performance analysis, scientific mapping, and network analysis—and not the underlying thematic discourses using co-citation analysis.

[3] Of the 3,998 articles in Scopus that deal with the broader topic of biodiversity finance, 10% were classified as "Economics, Econometrics and Finance" and 6% as "Business, Management and Accounting". Notably, 57% were classified as "Environmental Science" and 40% as "Agricultural and Biological Sciences". Publications can be assigned to several subject areas in Scopus.

Building on the growing but still fragmented body of finance-oriented research on biodiversity (Garel et al., 2024; Hunjra, 2025; Poyser, 2026) and on early bibliometric efforts in the field (Hutchinson and Lucey, 2024; Shehzad and Khan, 2024), this study is guided by three research questions: (i) What are the dominant thematic research streams in biodiversity finance, and how have they evolved over time? (ii) Which foundational publications and journals structure each of these streams? (iii) To what extent do these streams exchange knowledge, and where do the most pronounced silos lie? Addressing these questions, the study makes three contributions. First, it provides the largest co-citation-based thematic mapping of biodiversity finance to date, based on 3,998 articles and 189,456 references, extending the scope and method of prior bibliometric work that has so far relied on smaller corpora or purely descriptive performance and network analyses. Second, it identifies and characterizes eight research streams whose co-existence and partial tensions (market-based mechanisms vs. political-economy critiques; corporate reporting vs. community-based conservation) have not previously been mapped in an integrated way. Third, it quantifies limited cross-stream information exchange using density scores, providing an empirical diagnosis of fragmentation and a concrete basis for a focused research agenda and implications for policymakers, financial institutions, and corporate actors.

The paper is structured as follows: Section 2 outlines the data and methods used for the bibliometric analysis, while Section 3 presents the results, including a detailed review of the identified research streams, their prominent publications, and the evolution of the discourse. Section 4 discusses the findings, emphasizing the implications for policymakers and researchers, the study's limitations, and opportunities for future work. Section 5 concludes.

## 2  Data and Methods

### 2.1  Literature data and processing

In August 2024, the literature data were sourced from the Scopus database, which offers several advantages: Firstly, it allows searches to be limited to peer-reviewed articles, ensuring a standard level of quality. Secondly, Scopus enables systematic searches using specific terms. Lastly, it provides access to metadata, such as article references, which facilitates co-citation analysis. This analysis identifies relationships or similarities between publications by measuring the frequency with which two articles are cited together by a third party (Small, 1973). A high co-citation frequency indicates that both articles are significant within a particular discourse, as they form the basis for many authors' arguments and analyses. Therefore, co-citation clusters can be used to examine the intellectual structure of scientific discussions (White and Griffith, 1981).

We relied on the search string "*biodiversity\* AND financ\**"[4] across titles, abstracts, and keywords, yielding a sample of 3,998 scientific articles on biodiversity and finance; these papers contained 189,456 references. Upon initial inspection of the data, we identified that the references were not standardized, resulting in dozens of different strings that represented the same unique reference. To solve this challenge, we normalized references by removing punctuation and converting them to lowercase. Next, we applied MinHash Locality-Sensitive Hashing (LSH) to group similar references. LSH can be used to efficiently estimate the similarity between datasets, which is especially useful in handling large volumes of data (Jafari et al., 2021; Zamora et al., 2016). Furthermore, fuzzy matching was used to refine the grouping and identify records that are likely to be the same despite minor spelling or formatting differences (Chaudhuri et al., 2003). The refined literature sample was then used to calculate co-citation data.

---

[4] The use of the asterisks ensured that extended terms such as "*biodiversity-focused*," "*finance*," or "*financial*" are included in the sample.

A total of 412 references, each cited more than nine times, served as the foundation for our co-citation analysis. Following standard bibliometric procedures (Small, 2006), we applied a threshold value determined by an elbow criterion in the data distribution to simplify the interpretation process (Zupic and Čater, 2015). Bibliometric data tend to be highly skewed, with a small number of publications on a given topic contributing disproportionately to the scientific output and theoretical developments (Chen and Leimkuhler, 1986). Thus, for our analysis, it is sufficient to focus on the articles with the most co-citations. A co-citation matrix comprising 412 rows and columns—one for each publication—was constructed, with the cells indicating the frequency of co-citation between each pair of articles. This matrix was then employed for subsequent factor and network analyses.

## 2.2 Empirical strategy

Exploratory factor analysis is used to reduce the number of variables in the dataset, thereby simplifying its complexity without the need to predefine specific questions or hypotheses (Gorsuch, 1988). Applied to the co-citation matrix, this method groups articles into factors that, in our context, represent research streams. A research stream, therefore, consists of publications that share similar co-citation patterns (Ante, 2024). For statistical reasons, articles with low communalities were omitted from the sample, resulting in a total of 389 articles that served as the foundation for the factor analysis (Hogarty et al., 2005)—an adequate sample size (Comrey and Lee, 2013). The analysis produces two key metrics. First, factor loadings, which represent correlations between an article and a factor, range from -1 to 1 after Varimax rotation and Kaiser normalization (Gorsuch, 1988), indicating the degree of an article's alignment with a research stream. A loading of 0.4 typically indicates that an article belongs to a research stream, while values above 0.7 suggest a significant contribution (McCain, 1990). Thus, publications with loadings of at least 0.4 are recognized as part of a research stream. Nonetheless, the designation of these quantitatively derived streams necessitates a subjective assessment informed by a qualitative examination of the publications (Zupic and Čater, 2015). Secondly, factor scores, derived through regression analysis, rank individual articles within research streams. A higher factor score indicates greater relevance to the research stream (Gorsuch, 1988). In contrast to factor loadings, which indicate the degree to which an article aligns with a particular stream, factor scores emphasize an article's specific contribution to that stream. A review article may thus be appropriate in context (high factor loading) yet offer minimal contribution (low factor score).

Following the identification of distinct research streams, network analysis is used to investigate information flow among them. To achieve this, density scores are derived from the co-citation matrix, which illustrates intellectual exchange both between and within the various research streams. The UCINET software (Borgatti et al., 2002) is utilized. The actual co-citation relationships are compared with all possible relationships by converting the co-citation matrix to a binary format, in which cells are assigned a value of 1 to indicate the presence of a relationship and 0 to indicate its absence. Articles are then grouped according to the research streams identified in the factor analysis, resulting in an $8 \times 8$ matrix. This matrix quantifies the extent of information exchange among various research streams and within each stream. A value of zero signifies no information exchange, while a value of one denotes complete information exchange.

The combination of co-citation-based exploratory factor analysis (EFA) and network density analysis is intentionally chosen to match our research questions. EFA on a co-citation matrix is particularly well suited to recover the intellectual foundations of a field by grouping together publications that have historically been cited jointly, and is therefore well aligned with our first research question on thematic structure and evolution (Small, 1973; White and Griffith, 1981; Zupic and Čater, 2015). Alternative bibliometric techniques address different questions: bibliographic coupling captures short-term frontier similarities among recent papers rather than long-run intellectual structure; keyword co-occurrence reflects lexical rather than intellectual proximity; and topic modelling approaches such as LDA or BERTopic summarize abstract-level content but are agnostic to how the community itself organizes its knowledge through citations. Network-based density scores complement the EFA by quantifying cross-stream knowledge exchange, directly addressing our third research question and providing an empirical diagnosis of fragmentation that the aforementioned alternatives do not deliver.

## 3   Research streams in biodiversity finance

Table 1 provides an overview of the eight research streams identified by factor analysis, sorted by the proportion of variance explained and the share of the 389 articles in each stream. The general content of each stream is briefly summarized, along with the peak publication period, the principal academic journals, and the most influential publications. The factor analysis is performed using principal component analysis with varimax rotation and Kaiser normalization. A Kaiser-Meyer-Olkin (KMO) measure of sampling adequacy of 0.543 and a highly significant Bartlett test of sphericity ($p < 0.0001$) reveal that exploratory factor analysis is a suitable method of analysis (Gorsuch, 1988; Kaiser and Rice, 1974).

**Table 1. Research streams in biodiversity finance.**

| Name | Variance explained | Share of articles | Topic(s) | Peak periods | Principal journals | Formative publications |
|---|---|---|---|---|---|---|
| **Stream I:** Strategic and financial approaches in global biodiversity conservation | 10.1% | 15.2% | - **Conservation prioritization and cost efficiency**: Development of frameworks that integrate economic factors and decision theory to prioritize conservation efforts and enhance cost-effectiveness.<br>- **Global conservation planning and management costs**: Relevance of global and regional coordination, economies of scale in managing protected areas, and the variability of associated costs.<br>- **Financial sustainability and effectiveness of protected areas**: Sustainable financing mechanisms and assessing the effectiveness of protected areas in reducing biodiversity loss. | 2004 - 14 | Science (14%)<br><br>Nature (12%) | - Balmford et al. (2003)<br>- Waldron et al. (2013)<br>- Margules & Pressey (2000) |
| **Stream 2:** Impact and implementation of payments for environmental services (PES) in developing countries | 8.1% | 9.5% | - **Hydrological and environmental effects:** The impact of land use changes, particularly deforestation, on hydrological functions, including alterations in rainfall patterns, water yield, and erosion under different geographic and climatic conditions.<br>- **Institutional and market dynamics of PES:** The role of institutional settings, governance, stakeholder engagement, and clear land rights in the effectiveness of PES.<br>- **Economic aspects and resilience of environmental markets:** Analysis of environmental markets, especially carbon trading, their economic resilience and capacity to maintain value and transaction volumes. | 2008 - 10 | Ecological Economics (35%) | - Engel et al. (2008)<br>- Muradian et al. (2010)<br>- Wunder (2005) |
| **Stream 3:** Neoliberal influences and implications in environmental conservation | 5.6% | 8.7% | - **Neoliberalization of nature:** How neoliberal policies facilitate the commercialization and deregulation of the environment, focusing on privatization and market-driven strategies.<br>- **Capitalism and conservation integration:** The role of global capitalism in driving biodiversity conservation; how conservation efforts serve capitalist expansion, linking leisure and consumer habits directly to environmental impacts.<br>- **Critiques and alternatives to neoliberal conservation:** Critical views on neoliberal conservation methods, discussing how they commodify nature under the guise of protection and sustainability, and alternative approaches that decouple conservation from capitalist gains. | 2010 - 12 | Antipode (15%) | - Buscher et al. (2012)<br>- Igoe et al. (2010)<br>- Brockington et al. (2010) |
| **Stream 4:** Biodiversity offsets and conservation: policies and practicalities | 4.9% | 8.0% | - **Biodiversity offsets and policy development:** Development and comparison of biodiversity offset frameworks across various countries, detailing mechanisms to mitigate environmental impacts through compensatory conservation measures, which are implemented to achieve 'no net loss' or 'net gain' of biodiversity.<br>- **Efficacy and challenges of offset policies:** Effectiveness of biodiversity offsetting as a policy tool, discussing time lags, unpredictability of ecological restoration outcomes, and the complex trade-offs between development pressures and conservation goals.<br>- **Critical perspectives on market-based conservation:** Biodiversity offsets are shortfalls due to the complexity of ecological metrics, the difficulty in ensuring equivalent biodiversity gains, and potential to inadvertently encourage detrimental environmental practices. | 2010 - 15 | Biological Conservation (16%)<br><br>Conservation Letters (13%) | - Maron et al. (2012)<br>- Bull et al. (2013)<br>- McKenney et al. (2011) |
| **Stream 5:** Ecosystem services and biodiversity | 3.6% | 9.0% | - **Value and impact of ecosystem services:** Economic value of natural ecosystems which provide services like water purification, climate regulation, and biodiversity conservation.<br>- **Integration challenges in policy and management:** Challenges in integrating ecosystem services into effective landscape planning, management, and decision-making. Challenges include measuring, valuing, and systematically incorporating ecosystem service data.<br>- **Conservation strategies and biodiversity:** Analysis of conservation approaches that focus on ecosystem services and how they need to be effectively aligned with biodiversity goals to prevent potential trade-offs in conservation outcomes. | 2005 -10 | Science (11%)<br><br>Nature; Ecological Economics; PLoS Biology (9% each) | - Costanza et al. (1997)<br>- De Groot et al. (2002)<br>- Fisher et al. (2009) |

| Stream | | | | | | | |
|---|---|---|---|---|---|---|---|
| **Stream 6:** Integrating conservation and community interests in biodiversity management | 3.4% | 6.7% | - **Ecotourism as a conservation tool:** Effectiveness and economic incentives of ecotourism, questioning its viability as a sustainable conservation strategy and its real benefits to local communities and biodiversity.<br>- **Community-based conservation efforts:** Challenges and successes of engaging local communities in conservation efforts, highlighting the necessity for alignment between conservation objectives and local livelihoods to ensure mutual benefits and sustainable outcomes.<br>- **Effectiveness and impact of protected areas:** The role of protected areas in biodiversity conservation, assessing their effectiveness in achieving conservation goals and their socio-economic impacts on adjacent communities. | 2004 - 06 | Science (19%)<br><br>Trends in Ecology and Evolution (11%) | - Myers et al. (2000)<br>- Berkes (2004)<br>- Bruner et al. (2001) |
| **Stream 7:** Balancing agricultural intensification with biodiversity conservation | 2.6% | 6.2% | - **Agricultural practices and biodiversity:** The tension between increasing agricultural productivity and maintaining biodiversity, with focus on the ecological impacts of intensive farming practices and the potential benefits of habitat heterogeneity and agri-environment schemes.<br>- **Conservation value of traditional farming:** Ecological benefits of traditional farming systems, emphasizing their role in sustaining diverse species and offering guidelines for integrating biodiversity conservation into modern agricultural practices.<br>- **Effectiveness of agri-environment schemes:** The role of agri-environment schemes for enhancing biodiversity, suggesting mixed results and the need for improved design and implementation to truly benefit wildlife and ecosystem services. | 2000 - 03 | Journal of Applied Ecology (17%)<br><br>Science (13%) | - Kleijn and Sutherland (2003)<br>- Benton et al. (2003)<br>- Tscharntke et al. (2005) |
| **Stream 8:** Advancing the biodiversity and sustainability discourse through global and corporate reporting | 2.5% | 5.7% | - **Global sustainability reporting:** Global commitments towards sustainable development, integrating biodiversity conservation into broad socio-economic goals and tracking progress through transparent, accountable reporting mechanisms.<br>- **Corporate biodiversity reporting:** The extent and effectiveness of corporate biodiversity reporting, identifying gaps in current practices and suggesting that true sustainability accounting requires a comprehensive approach to capture impacts on biodiversity and ensure corporate accountability.<br>- **Emerging trends in environmental accountability:** Transformative approaches in biodiversity accountability, including integrated extinction accounting and biodiversity safeguards for global financial initiatives; a shift towards more rigorous, science-based, and ethical reporting practices to influence policy and corporate behavior towards biodiversity conservation. | 2015 - 21 | Accounting, Auditing and Accountability Journal (18%)<br><br>Business Strategy and the Environment (14%) | - Bongaarts (2019)<br>- Jones and Solomon (2013)<br>- Atkins and Maroun (2018) |

N = 389; Kaiser-Meyer-Olkin measure: 0.543; Bartlett's test: p < 0.0001.

The temporal development of research streams on biodiversity finance, visualized in Figure 1, reveals a shift in focus over the years. From 1948 to 1999, the majority of research (51.4%) fell into the "other publications" category, indicating a less structured approach. From 2000 to 2010, research diversified substantially, with an increased focus on strategic and financial approaches (Stream 1), PES implementation (Stream 2), and community interests in biodiversity (Stream 6), reflecting a broadening of specific themes.

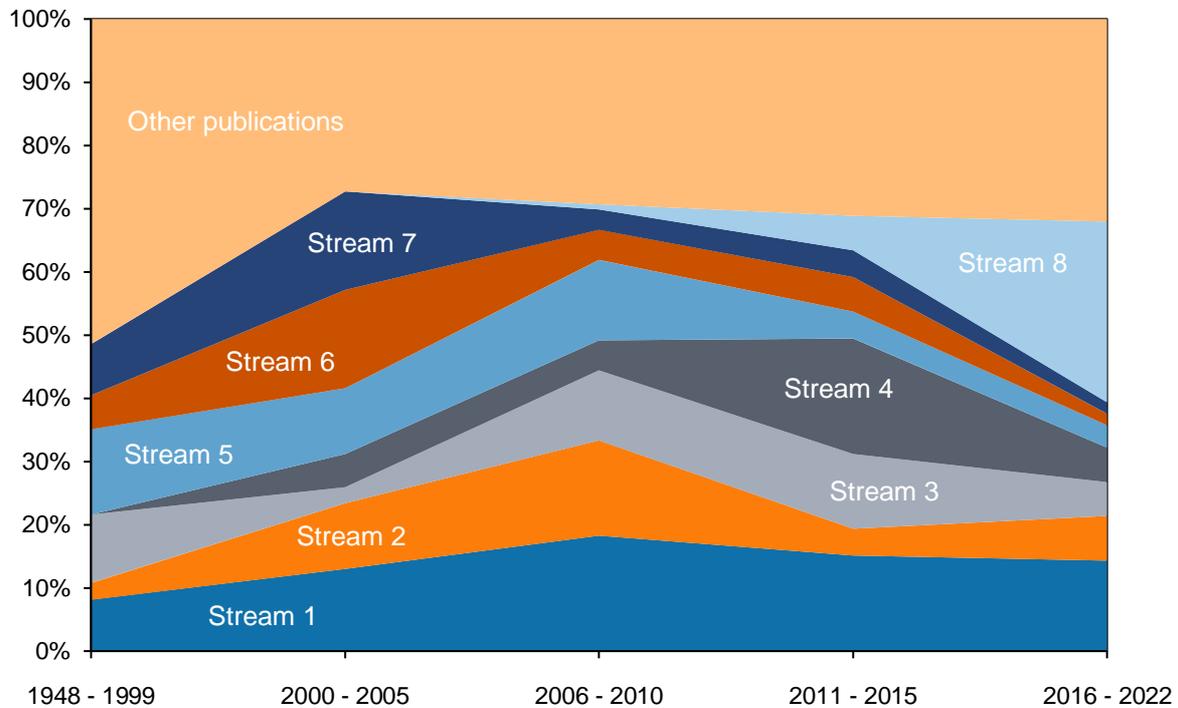

**Figure 1. Temporal development of research streams on biodiversity finance.**

Between 2011 and 2015, biodiversity offsets and conservation (Stream 4) became the dominant research area, suggesting a policy-driven shift in biodiversity finance. From 2016 to 2022, research on biodiversity and sustainability reporting (Stream 8, 28.6%) increased, while interest in earlier themes such as agricultural intensification (Stream 7) and community integration (Stream 6) declined substantially. Overall, the focus has moved towards sustainability reporting and structured financial approaches in biodiversity conservation.

### 3.1 Stream 1: Strategic and financial approaches in global biodiversity conservation

The first research stream comprises 58 publications (15.2% of the sample) and explains 10.1% of the variance. Table 2 shows the ten publications with the highest factor loadings and four others with high factor scores.

**Table 2. Articles on strategic and financial approaches in global biodiversity conservation**

| Reference | Factor loading | Factor score |
|---|---|---|
| Balmford et al. (2003) | 0.802 | 6.376 |
| Waldron et al. (2013) | 0.790 | 3.012 |
| Hickey and Pimm (2011) | 0.790 | 5.198 |
| Bruner et al. (2004) | 0.777 | 4.499 |
| Brockington and Scholfield (2010) | 0.773 | 3.065 |
| James et al. (2001) | 0.767 | 4.647 |
| McCarthy et al. (2012) | 0.738 | 3.835 |
| Armsworth et al. (2011) | 0.734 | 1.227 |
| Miller et al. (2013) | 0.717 | 3.437 |
| Bovarnick et al. (2010) | 0.716 | 2.188 |
| ... | ... | ... |
| Margules and Pressey (2000) | 0.629 | 5.074 |
| Naidoo et al. (2006) | 0.610 | 4.684 |
| Ferraro and Pattanayak (2006) | 0.563 | 4.059 |
| Watson et al. (2014) | 0.575 | 3.488 |

The challenge of financing biodiversity conservation is a persistent theme in the research stream, with greater emphasis given to understanding the variability in costs and the widespread underfunding of conservation efforts globally. Balmford et al. (2003) and Waldron et al. (2013) highlight the stark disparities in conservation funding, particularly in less developed regions where the benefit-to-cost ratio is highest. Despite this, these regions are often the most underfunded, leading to a critical shortfall in resources that hampers effective conservation. This issue is compounded by the findings of Bruner et al. (2004), who estimate that addressing the funding shortfall for managing existing protected areas in developing countries would require an additional $1 to $1.7 billion annually, with even more larger sums needed to expand protected areas and meet global biodiversity targets (James et al., 2001; McCarthy et al., 2012).

Given the substantial financial requirements, several studies have focused on the allocation and efficiency of conservation funding. Hickey and Pimm (2011) reveal that the World Bank's conservation investments are not always aligned with global biodiversity priorities, often favoring wealthier nations despite poorer countries needing more support. This pattern of uneven distribution is also evident in the work of Brockington and Scholfield (2010), who observe that conservation NGO funding in sub-Saharan Africa is concentrated in few organizations and regions, resulting in substantial underfunding in various areas, particularly in West Africa. To address such inefficiencies, Murdoch et al. (2007) and Wilson et al. (2006) suggest frameworks such as return on investment (ROI) for conservation planning, illustrating that considerable savings and enhanced outcomes can be realized when economic costs are evaluated alongside biological priorities. Margules and Pressey (2000) argue for a systematic approach to conservation planning that goes beyond the traditional reliance on reserves alone. They emphasize the need for strategies that manage entire landscapes, integrating both production and protection areas to effectively represent and preserve biodiversity. This aligns with the goals of more recent studies emphasizing the need for conservation plans to incorporate economic costs, leading to more efficient strategies that maximize biodiversity gains per dollar spent (Balmford and Whitten, 2003; Naidoo et al., 2006). This approach is further supported by Kark et al. (2009), who argue that multinational coordination can enhance conservation efficiency, particularly in complex regions like the Mediterranean Basin. However, as Larson et al. (2016) and Hein et al. (2013) note, the constraints of current funding mechanisms, such as philanthropy and PES highlight the need for innovative and diversified financial strategies to sustain long-term conservation efforts.

The governance and management of protected areas also play a critical role in the effectiveness of conservation funding. Miller et al. (2013) and Armsworth et al. (2011) emphasize the importance of adequate resourcing and governance structures that ensure funds are used effectively. Despite the expansion of protected areas globally, Coad et al. (2019) find that many sites require more resources for effective management, indicating that merely increasing the number of protected areas is insufficient without a concurrent emphasis on resource allocation and management quality. Finally, the literature stresses the importance of empirical evaluations to assess the effectiveness of conservation investments. Ferraro and Pattanayak (2006) advocate for evidence-based approaches to ensure that conservation funding leads to meaningful outcomes. Their research highlights the need for more effectively structured programs and assessments that may provide explicit direction on the optimal use of constrained conservation resources.

### 3.2 Stream 2: Impact and implementation of payments for environmental services (PES) in developing countries

The second research stream comprises 36 publications (9.5% of the sample) and explains 8.1% of the variance. Table 3 shows the eleven publications with the highest factor loadings and four others with high factor scores.

**Table 3. Articles on the impact and implementation of PES in developing countries**

| Reference | Factor loading | Factor score |
|---|---|---|
| Wunder (2008) | 0.860 | 2.832 |
| Muradian et al. (2010) | 0.857 | 5.960 |
| Wunder (2005) | 0.825 | 5.822 |
| Vatn (2010) | 0.811 | 3.080 |
| Pagiola (2008) | 0.796 | 2.712 |
| Wunder (2007) | 0.792 | 3.739 |
| Pagiola et al. (2005) | 0.790 | 4.246 |
| Wünscher et al. (2008) | 0.784 | 1.610 |
| Pattanayak et al. (2010) | 0.781 | 2.616 |
| Milder et al. (2010) | 0.778 | 2.510 |
| Turpie et al. (2008) | 0.778 | 1.638 |
| ... | ... | ... |
| Engel et al. (2008) | 0.756 | 10.115 |
| Wunder et al. (2008) | 0.772 | 5.121 |
| Landell-Mills and Porras (2002) | 0.740 | 3.110 |
| Ferraro and Kiss (2002) | 0.592 | 2.925 |

PES has gained prominence as an economic tool to align environmental conservation with incentives for landowners. Wunder (2008, 2005) presented PES as a market-oriented strategy for conservation, in which landowners receive compensation for delivering ecosystem services, and stress the need to harmonize efficiency with social goals. A PES scheme can be described as a voluntary, conditional arrangement between at least one "seller" and one "buyer," in which the seller agrees to provide a specific environmental service or to adopt a land use expected to generate that service (Wunder, 2007). While PES schemes are primarily designed to meet environmental goals, their potential to alleviate poverty has been a critical focus. Pagiola et al. (2005) and Wunder (2008) explored the role of PES in benefiting poor landholders, particularly in Latin America, though they caution that PES is not inherently a poverty-alleviation tool.

From an institutional economics perspective, Coase's (1960) theory on the allocation of social costs and Ostrom's (2005) work on institutional diversity underpins the economic rationale for PES. Muradian et al. (2010) and Vatn (2010) provided insights into the reductionist market conceptualization of PES, contending that institutional and political factors frequently influence the resultant outcomes. The findings of these studies indicate that PES schemes often rely on the engagement of government or community entities rather than operating solely as market mechanisms, leading to implementation challenges. Engel et al. (2008) elaborated on PES design challenges, emphasizing the need for careful targeting and flexible program structures.

PES programs have varied in effectiveness, as shown by comparative analyses of government- and user-financed schemes. Wunder et al. (2008) found that user-financed programs are typically more efficient and better tailored to local conditions, while government-financed schemes often struggle with inefficiency and conflicting objectives. Pattanayak et al. (2010) conducted a review of the additionality of PES schemes in developing countries. Their findings indicate that although certain programs are effective at reducing deforestation, the overall impact is often constrained by weak institutional frameworks.

Furthermore, Costa Rica's PSA program, one of the most notable examples of PES, has been widely studied. Pagiola (2008) noted that while the PSA program has successfully protected watersheds, its efficiency could be improved through better payment targeting. Wünscher et al. (2008) and Milder et al. (2010) emphasized the importance of spatial targeting to maximize conservation benefits and efficiency.

Despite the potential outlined in the current body of literature, publications indicate that PES encounters several challenges. Landell-Mills and Porras (2002) and Ferraro and Kiss (2002) questioned its long-term sustainability without ongoing financial support, while Kosoy and Corbera (2010) critiqued the commodification of ecosystem services. Nonetheless, PES continues to serve as an important instrument for environmental conservation and rural advancement as long as its design takes into account the local context and harmonizes environmental objectives with social factors.

### 3.3 Stream 3: Neoliberal influences and implications in environmental conservation

The third research stream comprises 34 publications (8.7% of the sample) and explains 5.6% of the variance. Table 4 shows the fourteen publications with the highest factor loadings, which include the articles with the highest factor scores of the stream.

**Table 4. Articles on neoliberal influences and implications in environmental conservation**

| Reference | Factor loading | Factor score |
|---|---|---|
| Castree (2008) | 0.915 | 5.171 |
| Brockington and Duffy (2010) | 0.909 | 5.663 |
| Büscher et al. (2012) | 0.884 | 7.025 |
| Fletcher (2010) | 0.879 | 3.090 |
| Sullivan (2013) | 0.876 | 5.110 |
| MacDonald and Corson (2012) | 0.876 | 3.049 |
| Harvey (2005) | 0.863 | 2.877 |
| Arsel and Büscher (2012) | 0.861 | 4.505 |
| MacDonald (2010) | 0.860 | 4.475 |
| Igoe et al. (2010) | 0.859 | 6.123 |
| Brockington et al. (2012) | 0.854 | 3.077 |
| Büscher and Fletcher (2015) | 0.854 | 2.916 |
| Igoe and Brockington (2007) | 0.851 | 4.004 |
| McAfee (1999) | 0.847 | 2.646 |

The neoliberalization of nature, a process whereby natural resources are commodified and subjected to market mechanisms, has been widely critiqued in environmental scholarship. Castree (2008) identifies fundamental logics driving this transformation, such as deregulation, privatization, market expansion, and the commodification of nature. These dynamics reframe environmental governance, positioning conservation within capitalist frameworks and reshaping how ecosystems are valued and managed. This trend reflects a broader shift toward neoliberal conservation, where market-based mechanisms like PES become central tools. However, critics, including Fletcher (2010), argue that this market-driven approach prioritizes economic efficiency at the expense of social equity and justice, resulting in the marginalization of vulnerable communities.

Further exploring these dynamics, Brockington and Duffy (2010) highlight how conservation has become intertwined with global capitalism, with biodiversity protection serving as a driver for capital accumulation. This perspective is echoed by Büscher et al. (2012), who argue that neoliberal conservation reframes nature as a commodity to be traded, thus facilitating its integration into capitalist markets. This commodification is perhaps most evident in what Büscher and Fletcher (2015) call "Accumulation by Conservation," where conservation initiatives are repurposed to generate profit under the guise of environmental protection. This process is emblematic of broader financialization trends in conservation, as Sullivan (2013) explains, with environmental assets such as biodiversity and ecosystem services being transformed into financial products. The establishment of markets for nature, including carbon trading and biodiversity credits, exemplifies a pivotal transformation in which ecosystems are reconceptualized as commodities traded within global financial markets. The financialization of nature introduces new actors and mechanisms into the conservation landscape (Sullivan, 2013). Investors, banks, and financial institutions increasingly view conservation as a new frontier for capital accumulation, leading to practices like nature banking and nature derivatives. These developments raise concerns about the implications of aligning conservation with profit motives, as natural resources become speculative assets rather than public goods. MacDonald and Corson (2012) further explore this issue through the lens of The Economics of Ecosystems and Biodiversity (TEEB), which seeks to quantify the economic value of ecosystems. TEEB's approach reflects the broader neoliberal trend of reducing complex ecological systems to economic abstractions, allowing nature to be incorporated into global markets. While this approach may attract political and financial support for conservation efforts, it risks oversimplifying and commodifying nature, which could ultimately jeopardize long-term ecological sustainability. A major outcome of these neoliberal conservation practices is the phenomenon of "green grabbing," in which land and resources are seized under the pretext of environmental protection. As Fairhead et al. (2012) demonstrate, "green grabs" often displace local communities, alienating them from their land and resources. Similarly, Fletcher and Breitling (2012) critique PES programs in Costa Rica, noting that while these initiatives are presented as market-based conservation tools, they often rely heavily on state intervention. The dependence on government assistance calls into question the idea that PES schemes operate solely as neoliberal market mechanisms, highlighting the complexity of their practical implementation.

Neoliberal conservation's reconfiguration of environmental governance also extends to the roles of NGOs and private sector actors. Corson (2010) emphasizes that NGOs, once regarded as counterbalances to corporate influence, have progressively collaborated with businesses to influence conservation policies. This collaboration, particularly in the realm of U.S. foreign aid for biodiversity, reflects the broader neoliberal trend of blurring the boundaries between state, corporate, and civil society actors. These partnerships often serve to advance market-based conservation initiatives, further entrenching capitalist frameworks in environmental governance. Harvey (2005) contextualizes these developments within the broader history of neoliberalism, noting how market-driven policies have redefined state responsibilities, transferring the onus of environmental protection from public institutions to private entities and market mechanisms.

### 3.4 Stream 4: Biodiversity offsets and conservation: policies and practicalities

The fourth research stream comprises 30 publications (8.0% of the sample) and explains 4.9% of the variance. Table 5 shows the ten publications with the highest factor loadings and two others with high factor scores.

**Table 5. Articles on biodiversity offsets and conservation: policies and practicalities**

| Reference | Factor loading | Factor score |
|---|---|---|
| Moilanen et al. (2009) | 0.909 | 4.496 |
| Bekessy et al. (2010) | 0.907 | 3.166 |
| Gordon et al. (2015) | 0.894 | 3.216 |
| Walker et al. (2009) | 0.892 | 5.575 |
| Maron et al. (2015) | 0.888 | 3.215 |
| Quétier and Lavorel (2011) | 0.885 | 5.905 |
| Curran et al. (2014) | 0.880 | 3.118 |
| Bull et al. (2013) | 0.873 | 6.216 |
| Villarroya et al. (2014) | 0.871 | 2.678 |
| Maron et al. (2012) | 0.868 | 7.921 |
| ... | ... | ... |
| Gardner et al. (2013) | 0.857 | 3.033 |
| McKenney and Kiesecker (2010) | 0.854 | 6.200 |

Biodiversity offsetting, a practice that aims to compensate for habitat loss caused by development through conservation or restoration efforts elsewhere, has gained traction as a tool for balancing economic growth with ecological preservation. However, biodiversity offset schemes face a variety of challenges, including uncertainty, time lags (Curran et al., 2014; Moilanen et al., 2009), and risks that complicate the achievement of conservation goals (Maron et al., 2015). Moilanen et al. (2009) argue that simplistic calculations of offset ratios fail to account for these complexities and advocate for a framework that incorporates uncertainty and time discounting to ensure robust conservation outcomes. Curran et al. (2014) echo these concerns, presenting empirical evidence that restoration, a common offset strategy, frequently encounters considerable delays, with biodiversity recovery potentially requiring centuries. Such time lags, coupled with uncertainties in restoration success, make it difficult to ensure that biodiversity gains match losses. Maron et al. (2015) further critique the assumption that restoration can compensate for biodiversity loss, pointing to inherent risks and time delays in offset policies that may result in net biodiversity loss rather than "no net loss" (NNL).

A second prominent issue is the creation of detrimental incentives within offset policies. Gordon et al. (2015) explore how poorly designed offsets may unintentionally encourage further biodiversity decline by creating economic and political incentives to lower conservation standards. Walker et al. (2009) add that the complexity and high expenses associated with biodiversity trading contribute to the shortcomings in realizing notable conservation results. Spash (2015) goes further, critiquing the commodification of biodiversity, suggesting that such approaches risk turning conservation into a market exercise that prioritizes economic interests at the expense of ecological integrity.

The concept of achieving NNL, central to many offset policies, presents additional challenges. Gardner et al. (2013) note that current NNL policies often rely on assumptions that are not ecologically feasible, with biodiversity losses being inadequately compensated by offset gains. Bull et al. (2013) analyze these discrepancies between theory and practice, while Maron et al. (2015) discuss how offset policies in Australia have "locked in" biodiversity declines by using baselines that assume ongoing deterioration rather than improvement.

Governance and implementation frameworks also pose difficulties. Villarroya et al. (2014) review biodiversity offset policies in Latin America, identifying considerable gaps in governance structures that undermine policy effectiveness. Similarly, McKenney and Kiesecker (2010) highlight the need for robust legal and institutional frameworks to ensure that offset policies are both enforceable and effective.

Finally, Quétier and Lavorel (2011) examine the technical challenge of achieving ecological equivalence in offset schemes. They argue that existing methods often fail to account for time delays and uncertainties about restoration success, leading to offsets that do not genuinely compensate for biodiversity loss. Bullock et al. (2011) discuss the conflicts between biodiversity restoration and ecosystem service restoration, noting that these goals are not always compatible. Suding (2011) adds that, though evolving, restoration science faces many obstacles to achieving consistent and reliable outcomes.

## 3.5 Stream 5: Ecosystem services and biodiversity

The fifth research stream comprises 35 publications (9.0% of the sample) and explains 3.6% of the variance. Table 6 shows the ten publications with the highest factor loadings and four others with high factor scores.

**Table 6. Articles on ecosystem services and biodiversity**

| Reference | Factor loading | Factor score |
|---|---|---|
| Balvanera et al. (2006) | 0.800 | 3.285 |
| Fisher et al. (2009) | 0.775 | 3.195 |
| Naidoo et al. (2008) | 0.775 | 3.924 |
| Naidoo and Ricketts (2006) | 0.775 | 4.057 |
| Boyd and Banzhaf (2007) | 0.736 | 3.599 |
| de Groot et al. (2002) | 0.726 | 4.283 |
| Chan et al. (2006) | 0.725 | 3.494 |
| Goldman et al. (2008) | 0.705 | 2.728 |
| Mace et al. (2012) | 0.698 | 2.159 |
| Hooper et al. (2005) | 0.694 | 1.997 |
| ... | ... | ... |
| Costanza et al. (1997) | 0.677 | 3.351 |
| Millennium Ecosystem Assessment (2005) | 0.651 | 8.040 |
| Sukhdev et al. (2010) | 0.595 | 6.645 |
| de Groot et al. (2010) | 0.593 | 3.972 |

Biodiversity is essential to the functioning of ecosystems and is intricately linked to fundamental ecosystem services, including carbon sequestration and soil-related services (Mace et al., 2012). Balvanera et al. (2006) show that higher biodiversity enhances ecosystem resilience and productivity, making it essential for maintaining ecosystem stability. Similarly, Hooper et al. (2005) argue that biodiversity notably impacts ecosystem services, contributing to their stability over time. Further, biodiversity both regulates ecosystem processes and is also a valuable resource that can be assessed and managed (Mace et al. 2012). These findings highlight the close link between biodiversity loss and the diminished ability of ecosystems to provide essential services, making biodiversity conservation crucial for ecosystem health.

Costanza et al. (1997) estimate that ecosystem services contribute trillions of dollars annually to human welfare, even though they are not traded in markets. This value highlights the importance of integrating ecosystem services into national and international policy frameworks (Sukhdev et al., 2010). De Groot et al. (2002) argue that incorporating ecosystem service valuation into landscape management enhances conservation strategies by accounting for ecological, social, and economic benefits. The economic valuation of ecosystem services has emerged as a key tool for integrating ecological considerations into decision-making. Fisher et al. (2009) stress the importance of clear definitions and classifications of ecosystem services to improve policy relevance and effectiveness. Boyd and Banzhaf (2007) advocate for standardized measurement units that align ecosystem services with traditional economic accounts. De Groot et al. (2002) contribute to the discourse by formulating a typology of ecosystem functions that enables consistent valuation across diverse regions and contexts. The relationship between conservation and ecosystem services offers greater opportunities for maximizing both ecological and economic benefits. Naidoo et al. (2008) map global conservation priorities and show that regions rich in biodiversity often overlap with areas that provide substantial ecosystem services. Naidoo and Ricketts (2006) demonstrate that the economic advantages of conservation, including carbon sequestration, frequently surpass the associated costs, thereby making conservation initiatives economically feasible. Chan et al. (2006) argue for integrating ecosystem services into conservation planning to maximize benefits for both biodiversity and human well-being. Goldman et al. (2008) show that ecosystem service projects attract more diverse funding and engage broader stakeholder support compared to traditional conservation efforts.

The Millennium Ecosystem Assessment (2005) highlights the critical need for sustainable management practices, as ecosystem degradation directly threatens human well-being by compromising essential services such as food security, water quality, and climate regulation. As ecosystems continue to degrade, aligning biodiversity conservation with ecosystem service protection becomes increasingly important to mitigate broader societal impacts. In summary, biodiversity underpins ecosystem functioning, which in turn supports ecosystem services crucial for human well-being. Recognizing the economic value of these services offers a practical path to addressing environmental degradation while supporting sustainable economic growth (Fisher et al., 2009; Naidoo et al., 2008). This integrated approach may ensure long-term environmental and societal benefits by aligning conservation with economic priorities.

### 3.6 Stream 6: Integrating conservation and community interests in biodiversity management

The sixth research stream comprises 26 publications (6.7% of the sample) and explains 3.4% of the variance. Table 7 shows the ten publications with the highest factor loadings and four others with high factor scores. It is important to note that no individual study indicates a factor loading exceeding 0.7. Furthermore, the study with the highest factor score, conducted by Myers et al. (2000), is categorized under research stream 1.

**Table 7. Articles on the integration of conservation and community interests in biodiversity management**

| Reference | Factor loading | Factor score |
| --- | --- | --- |
| Mittermeier et al. (2005) | 0.639 | 1.068 |
| Smith et al. (2003) | 0.564 | 1.478 |
| Lamb et al. (2005) | 0.543 | 0.704 |
| Bates et al. (2015) | 0.525 | 0.377 |
| Butchart et al. (2010) | 0.524 | 1.221 |
| Ribeiro et al. (2009) | 0.523 | 0.838 |
| Sodhi et al. (2004) | 0.487 | 0.798 |
| Krüger (2005) | 0.487 | 0.771 |
| Hansen et al. (2013) | 0.479 | 0.976 |
| Berkes (2004) | 0.478 | 3.224 |
| ... | ... | ... |
| Bruner et al. (2001) | 0.441 | 3.046 |
| Kiss (2004) | 0.431 | 2.124 |
| West et al. (2006) | 0.416 | 2.490 |
| Myers et al. (2000)* | 0.312 | 13.211 |

* The study is assigned to research stream 1 (factor loading = 0.463 for stream 1)

Biodiversity hotspots, such as Southeast Asia and the Brazilian Atlantic Forest, face urgent threats due to deforestation and habitat fragmentation. Mittermeier et al. (2005) and Ribeiro et al. (2009) highlight the critical need to prioritize conservation in these ecologically rich but vulnerable regions. In Southeast Asia, up to 42% of biodiversity could be lost by 2100 (Sodhi et al., 2004), while Hansen et al. (2013) document accelerating deforestation in tropical regions, particularly Indonesia and Malaysia. Both studies stress the need for targeted conservation strategies.

Protected areas remain central to biodiversity conservation. Naughton-Treves et al. (2005) report that while protected areas reduce deforestation, surrounding land use continues to isolate these areas. Bruner et al. (2001) and Chape et al. (2005) emphasize the importance of management activities such as enforcement and community involvement in enhancing the effectiveness of protected areas. However, Smith et al. (2003) show that poor governance in conservation priority areas often leads to inefficiencies and biodiversity loss.

The role of ecotourism in conservation remains a subject of debate. Krüger (2005) observes that, although ecotourism holds potential, its success is contingent upon sustainable planning and the active involvement of local communities. Lamb et al. (2005) call for large-scale reforestation to restore degraded tropical landscapes, benefiting both biodiversity and local communities. Schipper et al. (2008) further highlight the threat climate change poses to global biodiversity, particularly for marine and land mammals. In summary, addressing biodiversity loss requires integrated efforts across protected areas, governance reform, and restoration, particularly in tropical hotspots where socio-economic pressures and climate change compound the challenges.

### 3.7 Stream 7: Balancing agricultural intensification with biodiversity conservation

The seventh research stream comprises 24 publications (6.2% of the sample) and explains 2.6% of the variance. Table 8 shows the ten publications with the highest factor loadings and four others with high factor scores.

**Table 8. Articles on balancing agricultural intensification with biodiversity conservation**

| Reference | Factor loading | Factor score |
|---|---|---|
| Kleijn et al. (2006) | 0.804 | 3.232 |
| Benton et al. (2002) | 0.784 | 1.715 |
| Kleijn and Sutherland (2003) | 0.782 | 8.634 |
| Robinson and Sutherland (2002) | 0.766 | 3.175 |
| Benton et al. (2003) | 0.765 | 7.451 |
| Kleijn et al. (2001) | 0.742 | 3.507 |
| Tscharntke et al. (2005) | 0.720 | 6.891 |
| Vickery et al. (2001) | 0.697 | 1.583 |
| Batáry et al. (2015) | 0.672 | 1.560 |
| Green et al. (2005) | 0.632 | 3.201 |
| … | ... | ... |
| Phalan et al. (2011) | 0.559 | 3.166 |
| Tilman et al. (2002) | 0.615 | 3.092 |

Agricultural intensification has been a major driver of biodiversity decline as modern farming practices transform landscapes and ecosystems. Agri-environment schemes (AES) are designed to mitigate these impacts. Kleijn et al. (2006) and Kleijn and Sutherland (2003) found that while AES moderately benefits common species, they are less effective for endangered species, often due to weak study designs. Benton et al. (2003) emphasized that habitat heterogeneity within farmlands—such as hedgerows and field margins—is essential for enhancing biodiversity. Robinson and Sutherland (2002) highlighted that the loss of habitat diversity and insect populations has greatly affected farmland bird species in Britain.

The debate over land sharing versus land sparing offers two strategies for balancing agricultural productivity and biodiversity conservation. Green et al. (2005) suggest that high-yield farming could spare more land for conservation, though the success of either approach depends on species' adaptation to farming practices. Tscharntke et al. (2005) argue for a landscape perspective, emphasizing the need for both local and large-scale management to sustain biodiversity in agroecosystems. They suggest that sustainable intensification—achieving higher yields without harming the environment—is crucial, especially in smallholder-dominated regions.

Policy plays a critical role in this balance. Siebert et al. (2006)) found that financial incentives alone are insufficient to encourage biodiversity-friendly practices; social and cultural factors also matter. Wätzold and Schwerdtner (2005) argue that conservation policies must be cost-effective to maximize their impact. In summary, successful biodiversity conservation in agricultural landscapes requires a combination of well-designed AES, habitat diversity, sustainable farming practices, and effective policy frameworks. Balancing these elements is key to achieving both agricultural productivity and ecosystem health.

### 3.8 Stream 8: Advancing the biodiversity and sustainability discourse through global and corporate reporting

The eighth and final research stream comprises 22 publications (5.7% of the sample) and explains 2.5% of the variance. Table 9 shows the ten publications with the highest factor loadings and three others with high factor scores.

**Table 9. Articles on global and corporate biodiversity and sustainability reporting**

| Reference | Factor loading | Factor score |
|---|---|---|
| Addison et al. (2019) | 0.729 | 3.804 |
| Roberts et al. (2021) | 0.727 | 3.919 |
| Hassan et al. (2020) | 0.719 | 3.737 |
| Atkins and Maroun (2018) | 0.718 | 5.676 |
| Maroun and Atkins (2018) | 0.701 | 3.854 |
| Jones and Solomon (2013) | 0.699 | 5.880 |
| Gray (2010) | 0.679 | 2.994 |
| Boiral (2016) | 0.667 | 4.387 |
| Ceballos et al. (2017) | 0.667 | 2.105 |
| United Nations (2016) | 0.657 | 1.229 |
| … | ... | ... |
| Rimmel and Jonäll (2013) | 0.645 | 3.964 |
| van Liempd and Busch (2013) | 0.639 | 3.589 |
| Bongaarts (2019) | 0.624 | 6.773 |

The decline of biodiversity is an urgent global issue, increasingly gaining attention in the corporate sector. Addison et al. (2019) highlighted that although major corporations are acknowledging biodiversity, their commitments often need measurable targets and transparency. This gap between corporate statements and actions is also discussed by Roberts et al. (2021), who emphasize the growing need for systematic biodiversity accounting due to emerging scientific evidence linking human activities to environmental crises. Atkins and Maroun (2018) introduced the notion of "extinction accounting," advocating the incorporation of extinction prevention into corporate reporting. This approach aims to bolster corporate responsibility, ultimately fostering corporate transformation and mitigating species loss (Maroun and Atkins, 2018). Further, Jones and Solomon (2013) stress the need for interdisciplinary approaches to address biodiversity loss and promote corporate accountability effectively.

Factors driving corporate biodiversity disclosure were explored by Hassan et al. (2020), who found that external recognition, such as environmental awards and partnerships with biodiversity organizations, positively influences companies' willingness to report. Furthermore, their findings indicate that companies in high-biodiversity-risk sectors are more likely to engage in biodiversity reporting. This suggests that external motivations, such as stakeholder pressure and sector-specific risks, play a crucial role in influencing corporate behavior. More recent sector-specific evidence supports and extends this interpretation: analyzing biodiversity transparency in European banking, Raimo et al. (2026) document substantial heterogeneity in nature-related disclosures across institutions and link higher transparency to board composition and sustainability governance structures, highlighting that emerging frameworks such as the TNFD are beginning to reshape corporate disclosure practice in finance-intensive sectors.

However, challenges in biodiversity reporting remain substantial. Boiral (2016) explored impression management strategies in the mining sector, revealing that companies often use rhetorical techniques to downplay their environmental impacts, thereby maintaining legitimacy without demonstrating genuine accountability. Similar findings are reported by Rimmel and Jonäll (2013) and van Liempd and Busch (2013), who observed limited and inconsistent biodiversity disclosures by companies in Sweden and Denmark, often driven by superficial motivations rather than a genuine dedication to environmental stewardship.

Gray (2010) questioned whether current sustainability accounting practices truly address the broader concept of sustainability, highlighting contradictions and inconsistencies in corporate sustainability narratives. Ceballos et al. (2017) emphasized the severity of biodiversity loss, describing the ongoing sixth mass extinction as a result of human activity and its notable declines in vertebrate populations. The SDGs report (United Nations, 2016) and the global assessment report on biodiversity and ecosystem services (Bongaarts, 2019) further emphasize the global urgency of addressing biodiversity loss and implementing sustainable practices across all sectors.

In conclusion, while frameworks such as extinction accounting present potential solutions for improving corporate biodiversity accountability, many challenges remain. Addressing these issues requires a comprehensive approach involving interdisciplinary frameworks, external drivers, and genuine corporate commitment to biodiversity conservation. Businesses must move beyond impression management to deliver measurable outcomes that contribute to halting and reversing biodiversity loss.

### 3.9  Information exchange between research streams

Figure 2 presents a density score matrix, quantitatively representing the degree of information exchange between research streams in biodiversity finance. In theory, the scores range from 0 to 1, where 0 indicates no information exchange and 1 reflects complete exchange. Each cell represents a proxy to the level of shared concepts, methodologies, and knowledge flow between two research streams, underscoring the interconnectedness of the field.

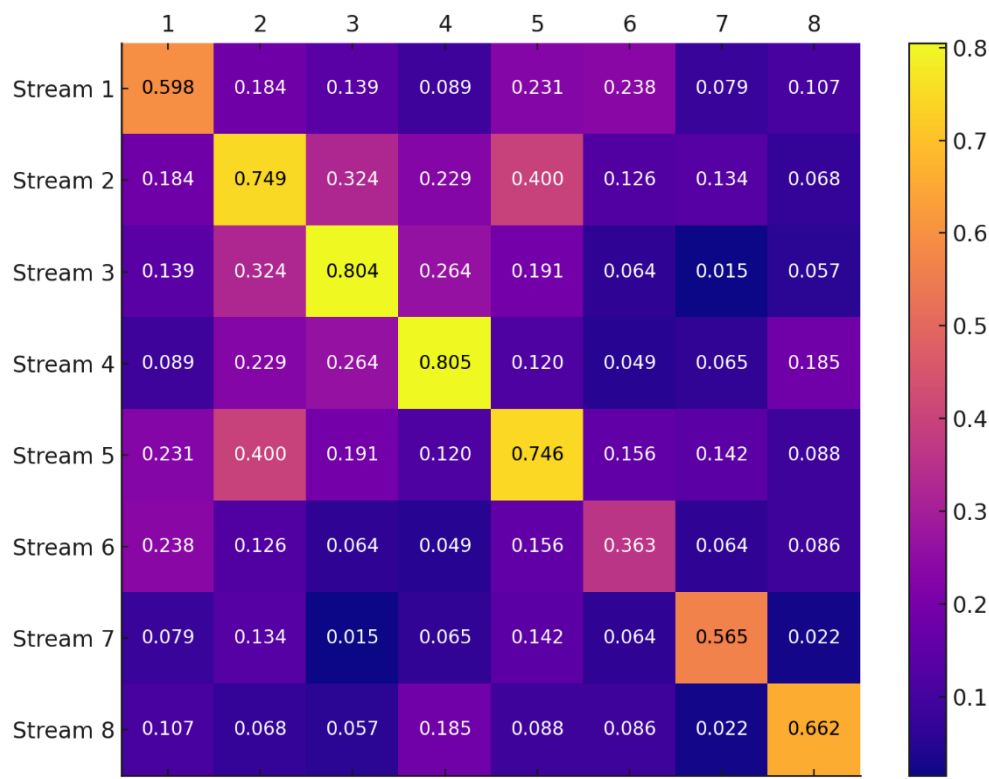

**Figure 2. Degree of information exchange between research streams (density scores)**

The most prominent interaction is between stream 1 and stream 5, with a score of 0.400, reflecting a shared focus on cost-effective conservation strategies and the valuation of ecosystem services. However, such strong connections are exceptions rather than the rule. Most inter-stream density scores are relatively low, pointing to distinct silos in how different areas of biodiversity conservation are approached. For example, stream 8 shows minimal exchange with streams focused on agriculture or neoliberal conservation, such as Stream 7, with a score of just 0.109.

The internal coherence within individual streams, reflected in the diagonal elements, tends to be higher than the inter-stream connections. For instance, streams 3 and 4 demonstrate strong internal focus, with scores of 0.804 and 0.805, respectively. This indicates that these streams maintain a concentrated research agenda but engage minimally with other areas—as indicated by low density scores with other research streams.

# 4 Discussion

## 4.1 Reflection on main results

The literature on biodiversity finance balances quantitative analyses, qualitative studies, and conceptual frameworks. From 1948 to 1999, most publications were not assigned to research streams, indicating a broad, less-structured approach to biodiversity finance. However, starting in the early 2000s, there was a greater diversification of research topics. This shift reflects an increasing recognition of the field's complexity and diversity.

A notable trend is the integration of economic principles into conservation strategies. Research streams 1 and 5 illustrate the growing focus on cost-efficiency, strategic prioritization, and the valuation of ecosystem services in biodiversity conservation. Within these streams, a substantial body of research reveals the pervasive challenge of inadequate funding for global nature conservation efforts (e.g., Balmford et al., 2003; Costanza et al., 1997). Research advocates for frameworks that aim to maximize biodiversity gains per unit of investment, aligning conservation outcomes with principles of economic rationality. This economic integration signifies a fundamental shift towards aligning conservation objectives with market mechanisms and financial incentives, thereby enhancing both effectiveness and sustainability. This approach may attract a broader range of stakeholders, particularly from the financial sector. However, it also raises critical questions regarding what is valued and what may be overlooked or disregarded in purely economic calculations.

In contrast, research streams 2 and 4 critically examine the role of market-based mechanisms, such as PES and biodiversity offsets. The literature highlights both the innovative potential and the challenges inherent in their implementation (e.g., Maron et al., 2012; Wunder, 2005). While these mechanisms offer novel solutions to financing conservation, they may come with unintended ecological consequences and concerns about equity and commodification of nature. Issues such as implementation complexities, reliance on government support, and the risk of privileging economic efficiency over ecological integrity highlight the need for careful design and robust governance structures. The implications are wide-reaching: while market-based mechanisms can mobilize essential financial resources, their success is heavily contingent on the socio-political context in which they are embedded, as well as the ability to safeguard against undermining core conservation goals.

Another critical pattern observed is the examination of neoliberal influences in environmental conservation, particularly within research Stream 3. Scholars critique how neoliberal policies facilitate the commodification of nature, thereby positioning conservation within capitalist frameworks (e.g., Büscher et al., 2012; Castree, 2008). Such an approach can marginalize vulnerable communities and prioritize economic gains over ecological integrity. This tension between adopting market mechanisms and ensuring social equity is a recurring theme, highlighting the need to balance financial incentives with broader social and ecological outcomes. The broader implication here is the ethical debate over whether conservation should be subject to market forces at all, as doing so may compromise long-term sustainability and justice for marginalized communities who rely on these ecosystems.

Despite advances in integrating economic approaches, research on the social dimensions of biodiversity finance remains less developed. Research streams 6 and 7, which focus respectively on community engagement and on balancing agricultural intensification with biodiversity conservation, are still in an embryonic phase. These areas emphasize the importance of aligning conservation efforts with community needs and ensuring sustainable agricultural practices, but few high-impact studies address these themes (e.g., Kleijn and Sutherland, 2003; Krüger, 2005). Moreover, the relatively low proportion of explained variance in these streams indicates that the social and cultural factors shaping conservation outcomes warrant a more comprehensive investigation. An emerging area of interest is corporate accountability and sustainability reporting in biodiversity conservation, highlighted in research Stream 8. Increasingly, major corporations are acknowledging their impact on biodiversity and incorporating biodiversity considerations into their sustainability frameworks (e.g., Addison et al., 2019; Jones and Solomon, 2013). Concepts such as "extinction accounting" (Atkins and Maroun, 2018) propose that corporations should integrate biodiversity metrics into their reporting practices to enhance transparency and accountability. However, the literature also highlights diverse challenges, including a lack of measurable targets, transparency issues, and the prevalence of impression management tactics. These issues imply that while corporate engagement is a promising development, there remains a substantial gap between rhetoric and action.

Overall, the density matrix reveals a broader picture of fragmentation in the biodiversity conservation research landscape. While individual streams are robust in their internal coherence, the low density of interstream connections suggests that interdisciplinary exchange remains underdeveloped. This fragmentation could pose challenges to addressing the inherently interconnected nature of biodiversity issues, which often require integrated approaches that bridge economic, social, and ecological perspectives. Greater collaboration and knowledge sharing between streams could enhance the collective impact of biodiversity research and support more comprehensive solutions to global conservation challenges.

In conclusion, the field of biodiversity finance demonstrates a dynamic interplay between economic integration, critical perspectives on market mechanisms, and emerging concerns regarding corporate accountability. The progress made in aligning conservation strategies with economic and financial principles is encouraging and points to a trend of increasing interdisciplinary collaboration. However, ongoing challenges—such as the ethical implications of commodifying nature, the underdeveloped social dimensions, and the necessity for robust governance structures—remain under-researched.

## 4.2 A two-dimensional synthesis of research streams

Taken together, the eight research streams are not merely a list of topics but can be organized along two cross-cutting dimensions that make the underlying tensions in biodiversity finance visible. The first dimension contrasts an economic-integration logic with a political-economy critique. Streams 1, 5, and 8 share the premise that conservation outcomes can be improved by making biodiversity legible to financial markets—through cost-efficient prioritization, ecosystem-service valuation, and corporate disclosure. Streams 2, 3, and 4 interrogate exactly this move: PES programs, biodiversity offsets, and neoliberal conservation are analyzed as sites where the commodification of nature produces power asymmetries, displaces local communities, and risks substituting accounting artifacts for ecological outcomes. The low inter-stream density scores between Stream 3 and Streams 1, 5, and 8 (see Figure 2) indicate that this tension is both conceptual and structural: the two camps rarely cite each other, which limits the scope for productive confrontation of their respective assumptions.

The second dimension distinguishes macro-level, global, and institutional perspectives (Streams 1, 3, 8) from micro-level, practice- and community-oriented perspectives (Streams 6, 7). The macro camp focuses on aggregate financing gaps, global conservation priorities, the logic of neoliberal governance, and corporate and sovereign reporting regimes; the micro camp engages with specific farming systems, community-based conservation arrangements, and ecotourism ventures where biodiversity outcomes are co-produced with local livelihoods. Streams 2 and 4 bridge these two poles, as PES and offset schemes are macro policy instruments whose success depends on micro-level implementation. Mapping the literature onto this 2×2 space clarifies where scholarly attention is concentrated (macro-economic and macro-critical work) and where it is thin (micro-level empirical finance research), and it exposes normative tensions—between efficiency and equity, between disclosure and accountability, between market expansion and community rights—that a purely descriptive cartography tends to obscure. A practical implication is that bridging research streams is not solely an intellectual virtue but a condition for designing biodiversity-finance instruments that are simultaneously scalable (macro), legitimate (critical), and effective on the ground (micro).

## 4.3 Towards a research agenda of biodiversity finance

Future research recommendations can be classified into multiple levels: some address the broader domain of current biodiversity finance studies, while others focus more narrowly on specific subfields, particularly as exemplified by distinct groups within the discipline.

**There is a need for more research on individual- and firm-level biodiversity finance.**

The increasing urgency of biodiversity conservation emphasizes the critical need for more research into biodiversity finance at both the individual and firm levels. While biodiversity loss has received substantial attention in policy and ecological research, the financial mechanisms that can support its preservation remain underexplored, particularly regarding how individuals and firms can contribute to sustainable solutions.

Research at the individual level could examine consumer preferences, willingness to pay for biodiversity conservation (Bhandari and Heshmati, 2010; Bhat and Sofi, 2021), and personal investment in biodiversity-related assets or projects (Garel et al., 2024). At an organizational level, research should examine business practices and motivations for investing in biodiversity protection (Nulkar and Bedarkar, 2020; Thompson, 2023). Organizations can influence outcomes by incorporating biodiversity factors into their financial decision-making processes, including corporate social responsibility (CSR) initiatives, environmental risk evaluations, or biodiversity offset programs. Research may examine how companies integrate biodiversity into sustainability reporting, their involvement in biodiversity financing via partnerships, or the impact of legislative frameworks and market incentives on corporate investments in biodiversity (Hassan et al., 2020). Ultimately, understanding the behavioral drivers influencing individual financial contributions to biodiversity initiatives is vital for designing effective financial instruments, such as biodiversity bonds, green investment funds, or conservation credits.

**There is a need to develop novel and innovative business models of biodiversity finance.**

There is a pressing need to develop innovative business models for biodiversity finance to address the growing challenges of effectively and sustainably funding conservation initiatives (Bishop et al., 2009). Traditional approaches to biodiversity funding, such as public grants or philanthropy, may need to be improved in scale, transparency, and reliability, necessitating the exploration of new models that can attract and mobilize private capital. Innovative business models could include mechanisms that link financial returns directly to conservation outcomes, incentivizing investment in biodiversity projects but further aligning conservation goals with business interests.

Additionally, hybrid models that integrate public, private, and philanthropic funding could provide more resilient and scalable financial solutions. For instance, public-private partnerships could help de-risk investments in biodiversity by leveraging government backing. In contrast, impact investment funds can attract socially conscious investors by promising measurable ecological and financial returns. To advance biodiversity finance, research is needed to identify, test, and refine these emerging business models. By focusing on innovative approaches that align economic incentives with conservation, the financial sector can play a pivotal role in funding biodiversity at the scale necessary to address the current environmental crisis.

**There is a need for increased cross-pollination of research between biodiversity finance subfields.**

Biodiversity finance research is fragmented, with various subfields addressing distinct aspects of funding and conservation strategies (see Figure 2). However, there is a pressing need for increased cross-pollination between these subfields to foster more comprehensive and integrated approaches.

Current research tends to isolate topics such as conservation finance, impact investing, corporate biodiversity management, and ecosystem services valuation. Contemporary research segregates subjects such as conservation finance, impact investing, corporate biodiversity management, and the assessment of ecosystem services. While each area provides unique insights, the lack of cross-sectoral interaction hinders the development of holistic solutions needed to address the multifaceted issue of biodiversity loss. Fullerton's (2015) study on regenerative capitalism and economies demonstrates the necessity of a comprehensive perspective in biodiversity management within a regenerative economy. This approach requires redefining wealth beyond financial metrics to include natural, living, cultural, social, and financial capitals, emphasizing their interconnectedness. Such a shift suggests that fostering enhanced collaboration between thematic and academic subfields could lead to innovative financial models and methods. For example, integrating insights from ecosystem service valuation with corporate biodiversity investments could improve the integration of natural capital in corporate decision-making processes. Moreover, interdisciplinary research that integrates conservation finance techniques with social impact investments may open new pathways to align financial returns with biodiversity outcomes. Promoting dialogue and joint research efforts across these subfields could deepen understanding of biodiversity finance and lead to more robust frameworks that can mobilize capital at scale for biodiversity conservation. Expanding this collaborative research agenda is crucial for creating synergies that strengthen the financial tools available to protect global biodiversity.

**There is a need to describe, consolidate, and optimize biodiversity finance-related data.**

To improve the quality and scope of analysis, there persists a need to describe, review, consolidate, and optimize biodiversity finance-related data (Costello et al., 2013; Costello and Wieczorek, 2014; Soberón and Peterson, 2004). Currently, biodiversity finance data is dispersed across multiple platforms and fields, including public financial records, environmental databases, corporate sustainability reports, and international conservation programs. These data sources vary widely in format, quality, and accessibility, making it difficult for researchers and practitioners to draw meaningful comparisons or conduct in-depth analyses. Describing the existing data landscape is a crucial first step, as it can reveal gaps, inconsistencies, and potential areas for standardization.

Consolidating this data into unified platforms or frameworks would enable more comprehensive insights into biodiversity finance flows, from public funding and private investments to conservation outcomes. Such consolidation could also facilitate better tracking of financial commitments, expenditures, and impacts, providing stakeholders with a clearer picture of how biodiversity finance is evolving globally. Optimization efforts are equally important and involve the development of standardized metrics, reporting guidelines, and data-sharing protocols. This would allow for a more consistent and transparent analysis of financial flows, enabling the identification of trends, best practices, and areas for intervention.

**There is a need to assess the role of technology in biodiversity finance.**

With the rapid acceleration of biodiversity loss, research should evaluate the role of technology in biodiversity financing, such as monitoring, reporting, and verification (MRV) systems. Technological breakthroughs have the potential to revolutionize the funding, monitoring, and management of biodiversity conservation; however, their effective integration into biodiversity finance remains to be investigated. Emerging technologies such as blockchain, artificial intelligence (AI), satellite imaging, and big data analytics offer new tools for enhancing transparency, efficiency, and impact in biodiversity finance. Blockchain, for instance, could facilitate more secure and transparent financial transactions related to biodiversity projects, such as tracking investments in conservation or enabling biodiversity credits and offset markets (Kochupillai et al., 2021; Stuit et al., 2022). AI and big data can improve risk assessments and impact evaluations by analyzing large-scale ecological and financial datasets (Bayraktarov et al., 2019; Hughes et al., 2024), helping investors make more informed decisions regarding biodiversity-related assets. Recent work further argues that the coupling of digital and green finance may act as a catalyst for biodiversity conservation by improving data infrastructures, lowering transaction costs, and enabling novel instruments for channelling capital toward nature-positive projects (Guo et al., 2026). Future research is needed to empirically test whether such digital–green finance synergies translate into measurable biodiversity outcomes or primarily improve the efficiency of existing instruments.

Furthermore, technology can improve the oversight of biodiversity results associated with financial investments. Remote sensing and satellite technology provide real-time monitoring of environmental changes (White et al., 2021), enhancing understanding of the effectiveness of conservation initiatives funded by biodiversity funding systems. These methods also have the potential to enhance accountability and ensure that financial transactions yield quantifiable biodiversity benefits. Research evaluating technology's applicability, efficiency, and constraints across many contexts is essential to fully harness its promise in biodiversity financing. Moreover, understanding how technological advancements can be optimally leveraged to enhance biodiversity finance will be crucial for unlocking new opportunities to scale conservation initiatives and attract broader financial participation.

**There is a need to engage more closely with the outcomes of biodiversity finance activities.**

While substantial resources are directed toward biodiversity finance, a gap remains between financial inputs and the tangible ecological and social outcomes they aim to achieve. Effective engagement requires both tracking financial flows and systematically assessing the real-world impacts of these investments on biodiversity. This involves measuring the extent to which biodiversity finance initiatives contribute to the restoration, protection, and sustainable management of ecosystems. By closely monitoring outcomes, researchers and policymakers can better understand which financial mechanisms and strategies are delivering the most substantial conservation benefits and where improvements are needed.

Furthermore, an increased emphasis on outcomes may facilitate the identification of unintended consequences, including financial investments that fail to support biodiversity conservation or, more critically, exacerbate environmental degradation. Addressing outcomes necessitates an examination of the wider social and economic implications of biodiversity finance. This includes the well-being and empowerment of local communities, the equitable allocation of resources, long-term sustainability, and the mitigation of technology- or crypto-driven colonialism. To foster this engagement, it is essential to establish comprehensive frameworks and regulatory-compliant methodologies that link financial investments to conservation outcomes. Thus, biodiversity finance has the potential to adopt a more results-oriented approach, enhancing its efficiency in addressing the global biodiversity crisis and ensuring that financial resources align with intended environmental outcomes.

# 5 Conclusion

Over recent years, biodiversity finance has become a pivotal focus within the broader discourse on environmental science and finance. While individual studies have assessed specific mechanisms and initiatives within biodiversity finance, a thorough evaluation of the field's interconnected and multidisciplinary development has been lacking. This review applies a blend of bibliometric techniques—co-citation and density analysis—to conduct a detailed thematic analysis of the biodiversity finance literature.

The study offers three novel contributions. First, the bibliometric results shed light on the complex web of relationships between scholars, key publications, and thematic clusters in biodiversity finance. This enables a more nuanced understanding of the intellectual structure underpinning the field. Secondly, we systematically delineate the subfields within biodiversity finance and analyze their characteristics and foundational works. The interdisciplinary nature of biodiversity finance, which encompasses economics, ecology, finance, and policy, makes this approach especially valuable for charting its academic terrain and identifying new pathways for exploration. Third, we outline a research agenda, emphasizing key areas where forthcoming studies could greatly enhance understanding of biodiversity finance.

Beyond these academic contributions, our findings also carry concrete implications for three audiences. For policymakers and regulators, the late emergence and internal coherence of an intellectual stream on global and corporate biodiversity reporting, together with its limited density with agricultural and community, suggests that disclosure frameworks such as the TNFD risk developing in relative isolation from the accumulated empirical knowledge on where biodiversity outcomes are actually produced. Aligning disclosure requirements with the evidence base on PES, offsets, and community-based conservation would help avoid a reporting regime that is technically sophisticated but disconnected from on-the-ground outcomes. For financial institutions and investors, the observed fragmentation implies that biodiversity-risk integration, product design (e.g., biodiversity bonds, blended finance), and due diligence currently rest on a narrow slice of the available evidence, predominantly from research streams 1, 5, and 8. A more systematic integration of insights from Streams 2, 3, and 4 would help institutions anticipate implementation risks, legitimacy concerns, and equity-related backlash that have repeatedly undermined market-based conservation instruments. For researchers, the density matrix itself serves as a diagnostic tool, identifying the pairs of streams in which cross-citation is weakest and therefore where interdisciplinary research is most likely to generate novel insights; the two-dimensional synthesis developed above provides a structured entry point for such work.

Despite the strengths of our methodology, there are inherent limitations. The bibliometric techniques employed, though systematic and transparent, necessitated specific decisions regarding sample selection, data refinement, and thematic area naming. These choices, by their nature, inevitably mean that certain peripheral developments or niche studies may not be captured. Consequently, while our analysis highlights the central trends in biodiversity finance, it may not account for pioneering or less prominent research on the field's margins. Specifically, the analysis is based solely on Scopus-indexed peer-reviewed articles and therefore does not cover books, book chapters, grey literature (including reports by the TNFD, the CBD, and the World Bank), or non-English-language scholarship; the data cut-off of August 2024 implies that the most recent wave of publications on nature-related financial disclosure, biodiversity-linked bonds, and AI-enabled MRV may be underrepresented.

## Declaration of generative AI and AI-assisted technologies in the writing process

During the preparation of this work, the authors used ChatGPT and Grammarly to improve the readability and language of the manuscript. After using the tools, the authors reviewed and edited the content as needed and take full responsibility for the content of the published article.